\documentclass[lettersize,journal]{IEEEtran}

\IEEEoverridecommandlockouts
\usepackage{amsmath}
\usepackage{amsthm,amssymb,amsfonts,mathrsfs,mathtools,amsbsy}
\usepackage{bm}
\usepackage{algorithmic}
\usepackage{algorithm}
\usepackage{array}
\usepackage[caption=false,font=normalsize]{subfig}
\usepackage{textcomp}
\usepackage{placeins}
\usepackage{url}
\usepackage{verbatim}
\usepackage{graphicx}
\usepackage{cite}
\usepackage[inline]{enumitem}
\usepackage{xspace}
\usepackage[table]{xcolor}
\usepackage{hyperref}

\colorlet{tableMulti}{red!20}
\colorlet{tableSingle}{cyan!30}

\newcommand{\IntegerP}{\mathbb{N}}
\newcommand{\IntegerPP}{\mathbb{N}_*}

\newcommand{\Complex}{\mathbb{C}}

\newcommand\vect[1]{\mathbf{#1}}
\newcommand\vectcal[1]{\mathbfcal{#1}}
\newcommand*{\hermconj}{\mathsf{H}}
\DeclarePairedDelimiter\norm{\lVert}{\rVert}
\DeclarePairedDelimiterX\innerp[2]{\langle}{\rangle}{#1
  \mathop{}\delimsize\vert\mathop{} #2}

\DeclarePairedDelimiterX\Set[1]{\lbrace}{\rbrace}%
{  #1 }

\DeclareMathAlphabet\mathbfcal{OMS}{cmsy}{b}{n}
\DeclareMathAlphabet\mathbfit{OML}{cmm}{b}{it}
\DeclareMathAlphabet\mathbfscr{OMS}{mdugm}{b}{n}

\DeclareMathOperator{\bdiag}{bdiag}

\makeatletter
\newcommand*{\ie}{%
  \@ifnextchar{,}%
  {i.e.}%
  {i.e.,\@\xspace}%
}
\newcommand*{\eg}{%
  \@ifnextchar{,}%
  {e.g.}%
  {e.g.,\@\xspace}%
}
\newcommand*{\cf}{%
  \@ifnextchar{,}%
  {cf.}%
  {cf.,\@\xspace}%
}
\makeatother

\usepackage{stackengine}
\stackMath
\newlength\matfield
\newlength\tmplength

\title{Multi-Linear Kernel Regression\\ and Imputation in Data Manifolds}

\author{Duc Thien Nguyen and Konstantinos Slavakis\IEEEauthorrefmark{1}%
  \thanks{\IEEEauthorrefmark{1}D.~T.~Nguyen and K.~Slavakis are with the Department of
    Information and Communications Engineering, Tokyo Institute of Technology, Tokyo, Japan
    (e-mails: \{nguyen.t.au, slavakis.k.aa\}@m.titech.ac.jp).}\vspace{-20pt}
}

\begin{document}
\maketitle

\begin{abstract}
  This paper introduces an efficient multi-linear nonparametric (kernel-based) approximation
  framework for data regression and imputation, and its application to dynamic
  magnetic-resonance imaging (dMRI). Data features are assumed to reside in or close to a
  smooth manifold embedded in a reproducing kernel Hilbert space. Landmark points are
  identified to describe concisely the point cloud of features by linear approximating patches
  which mimic the concept of tangent spaces to smooth manifolds. The multi-linear model effects
  dimensionality reduction, enables efficient computations, and extracts data patterns and
  their geometry without any training data or additional information. Numerical tests on dMRI
  data under severe under-sampling demonstrate remarkable improvements in efficiency and
  accuracy of the proposed approach over its predecessors, popular data modeling methods, as
  well as recent tensor-based and deep-image-prior schemes.
\end{abstract}

\begin{IEEEkeywords}
  Imputation, kernel, manifold, MRI, regression.\vspace{-10pt}
\end{IEEEkeywords}



\section{Introduction}

\IEEEPARstart{D}{ynamic} magnetic resonance imaging (dMRI) is a popular non-invasive imaging
modality for observing body organ movement, with rich potential in cardiac and neurological
diagnosis~\cite{zhi2000principles}. DMRI stands out as an application domain where regression
grapples with all of the archetypal data-analytic bottlenecks: large dimensionality due to the
image data, dynamic data patterns due to dMRI's time component, missing data due to
under-sampling, and strong but unknown spatio-temporal correlations since, often, dMRI monitors
structured movement; \eg, a beating heart~\cite{liang1994efficient}.

It comes, thus, as no surprise that numerous data-modeling approaches have been proposed for
regression and imputation on dMRI data: compressed sensing~\cite{lustig2007sparse,
  gamper2008compressed, liang2012k, feng2016xd}, low-rank models~\cite{otazo2010combination,
  lingala2011ktslr, zhao2012image}, and learning strategies based on
dictionaries~\cite{awate2012spatiotemporal, wang2014compressed, caballero2014dictionary,
  Wang2017parallel, ravishankar2017.lassi}, transforms~\cite{wen2017frist},
manifolds~\cite{Usman.Manifold.15, Chen:MA:IEEETMI:17, poddar2016dynamic, nakarmi2017m,
  nakarmi2018mls}, kernels~\cite{nakarmi2017kernel, poddar2019manifold, arif2019accelerated},
and tensors~\cite{liu2012tensor, Signoretto:arxiv:13, Kanagawa:GaussProcTensor:16,
  Li_Ye_Xu_2017}. Recent efforts on imputation-by-regression revolve also
around deep-learning (DeepL) approaches~\cite{schlemper2018deep, Biswas:MoDL-SToRM:19,
  liang2020deep, sandino2020compressed}, which rely on time-consuming processes to learn from
training data prior to reconstructing test data. Notwithstanding, concerns were raised
in~\cite{antun2020instabilities} via numerical tests which highlighted potential instabilities
of DeepL approaches. Motivated by deep image priors (DIP), DeepL networks have also been used
as implicit structural priors in regression for dMRI to avoid the use of training data and
potential over-fitting issues~\cite{Zou:TMI:21, yoo2021time}.

Departing from all of the previous schemes, this paper offers an extension of the novel
nonparametric data-modeling approach of~\cite{shetty2020bilmdm, slavakis2022krim}, coined
hereafter \textit{multi-linear kernel regression and imputation in data manifolds
  (MultiL-KRIM).} MultiL-KRIM introduces a multi-linear matrix decomposition in data modeling
to offer two-pronged innovation over its bi-linear predecessors KRIM~\cite{slavakis2022krim}
and BiLMDM~\cite{shetty2020bilmdm}:
\begin{enumerate*}[label=\textbf{(\roman*)}]

\item unlike KRIM and BiLMDM, where dimensionality-reduction \textit{pre-steps} are
  disassociated from the regression task, MultiL-KRIM connects dimensionality reduction
  directly with the regression task by enabling its inverse-problem solution to identify the
  ``optimal'' dimensionality-reduced rendition of a kernel matrix which contributes in minimum
  data-recovery error; and

\item it exploits its multiple matrix factors to promote efficient computations in its
  inverse-problem algorithmic solution.

\end{enumerate*}

MultiL-KRIM retains also the attributes which differentiate KRIM and BiLMDM from
state-of-the-art modeling approaches: unlike low-rank~\cite{otazo2010combination,
  lingala2011ktslr, zhao2012image, wen2017frist}, dictionary-learning
\cite{awate2012spatiotemporal, wang2014compressed, caballero2014dictionary, Wang2017parallel,
  ravishankar2017.lassi} and tensor~\cite{liu2012tensor, Signoretto:arxiv:13,
  Kanagawa:GaussProcTensor:16, Li_Ye_Xu_2017}
models, which promote a ``blind decomposition'' of the data matrix/tensor, MultiL-KRIM
incorporates the underlying data-manifold geometry directly into data representations, but not
via graph-Laplacian-matrix regularizers which are widely used in manifold-learning
approaches~\cite{Usman.Manifold.15, Chen:MA:IEEETMI:17, poddar2016dynamic}. MultiL-KRIM adopts
instead a ``collaborative-filtering'' modeling approach to identify ``optimal'' and
manifold-cognizant combinations of the observed data features for regression and
imputation. MultiL-KRIM needs no training data to operate, builds a nonparametric regression
estimate to reduce the dependence of its modeling assumptions on the probability distribution
of the data~\cite{Gyorfi:DistrFree:10}, and offers an explainable learning paradigm via simple
geometric arguments, unlike DeepL schemes, which are based, in general, on perplexed and
cascading non-linear function layers.

Numerical tests on synthetic dMRI data show that MultiL-KRIM outperforms several
state-of-the-art methods, including the total-variation tensor-based
scheme~\cite{Li_Ye_Xu_2017} and the deep-image-prior (DIP) propelled~\cite{yoo2021time}, while
matching at the same time the recovery-error performance of its predecessor
KRIM~\cite{shetty2020bilmdm, slavakis2022krim}, but with computational times which are lower
than and can reach down to one-third of those of KRIM.

\section{Data Collection and Formation in dMRI}

\begin{figure*}[ht]
  \centering
  \subfloat[The (k,t)-space \label{fig:dmri.ktspace}]
  {\includegraphics[height=100pt]{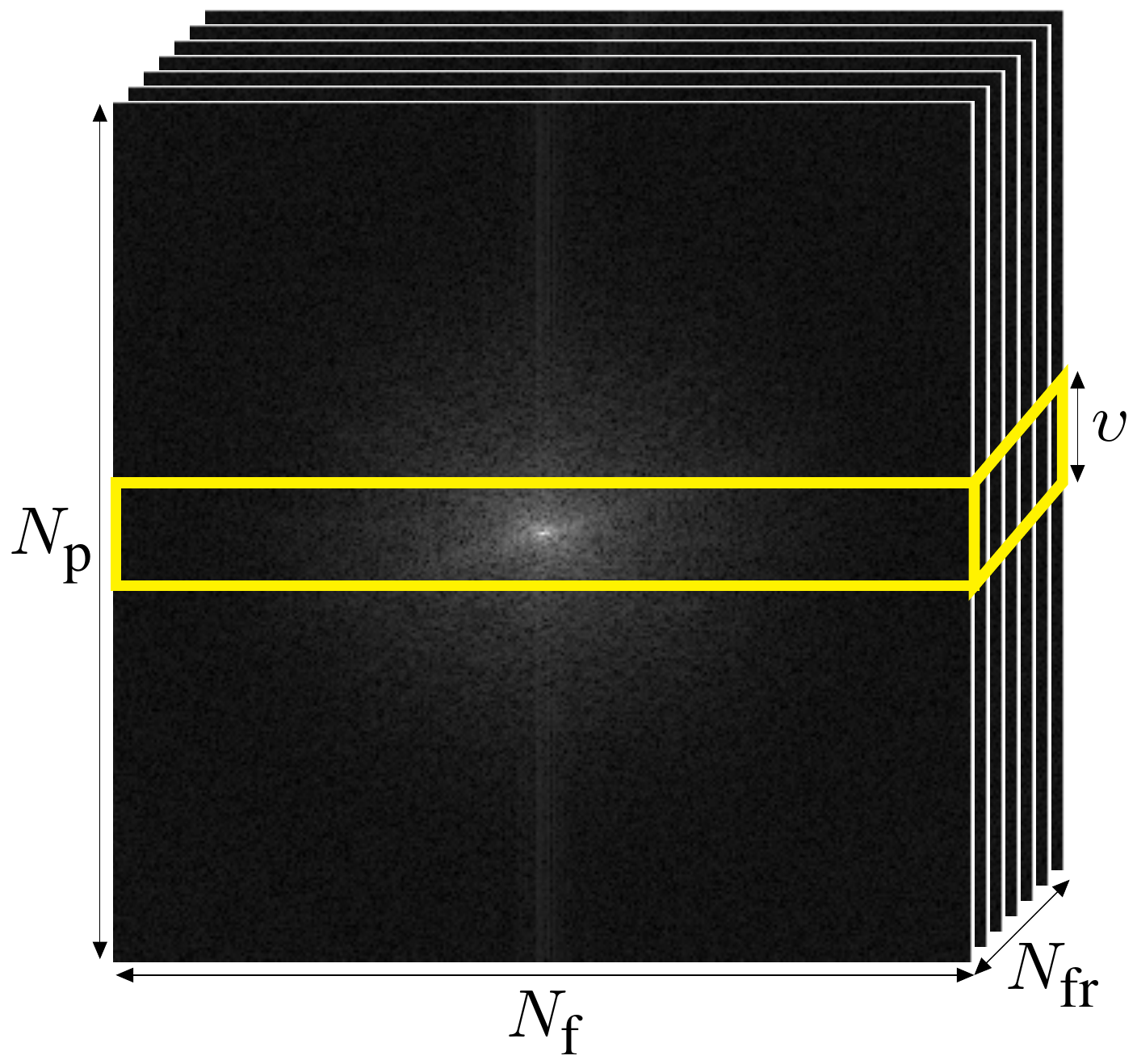}}\hspace{8pt}
  \subfloat[Cartesian sampling \label{fig:cartesian.sampling}]
  {\includegraphics[height=100pt]{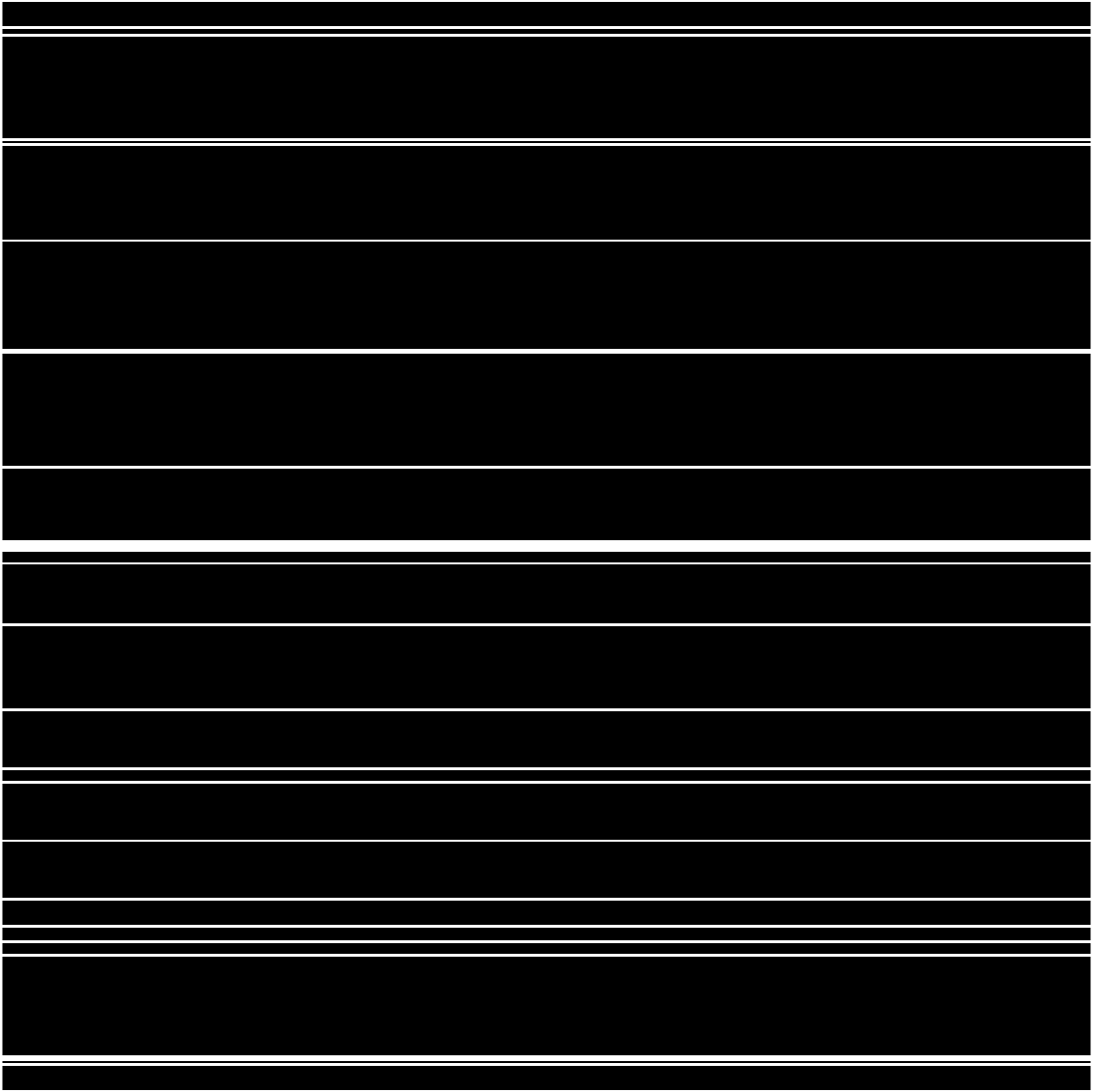}}\hspace{1pt}
  \subfloat[Radial sampling \label{fig:radial.sampling}]
  {\includegraphics[height=100pt]{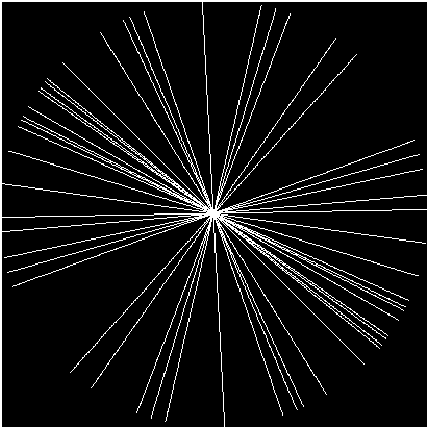}}\hspace{2pt}
  \subfloat[The image domain \label{fig:dmri.image}]
  {\includegraphics[height=100pt]{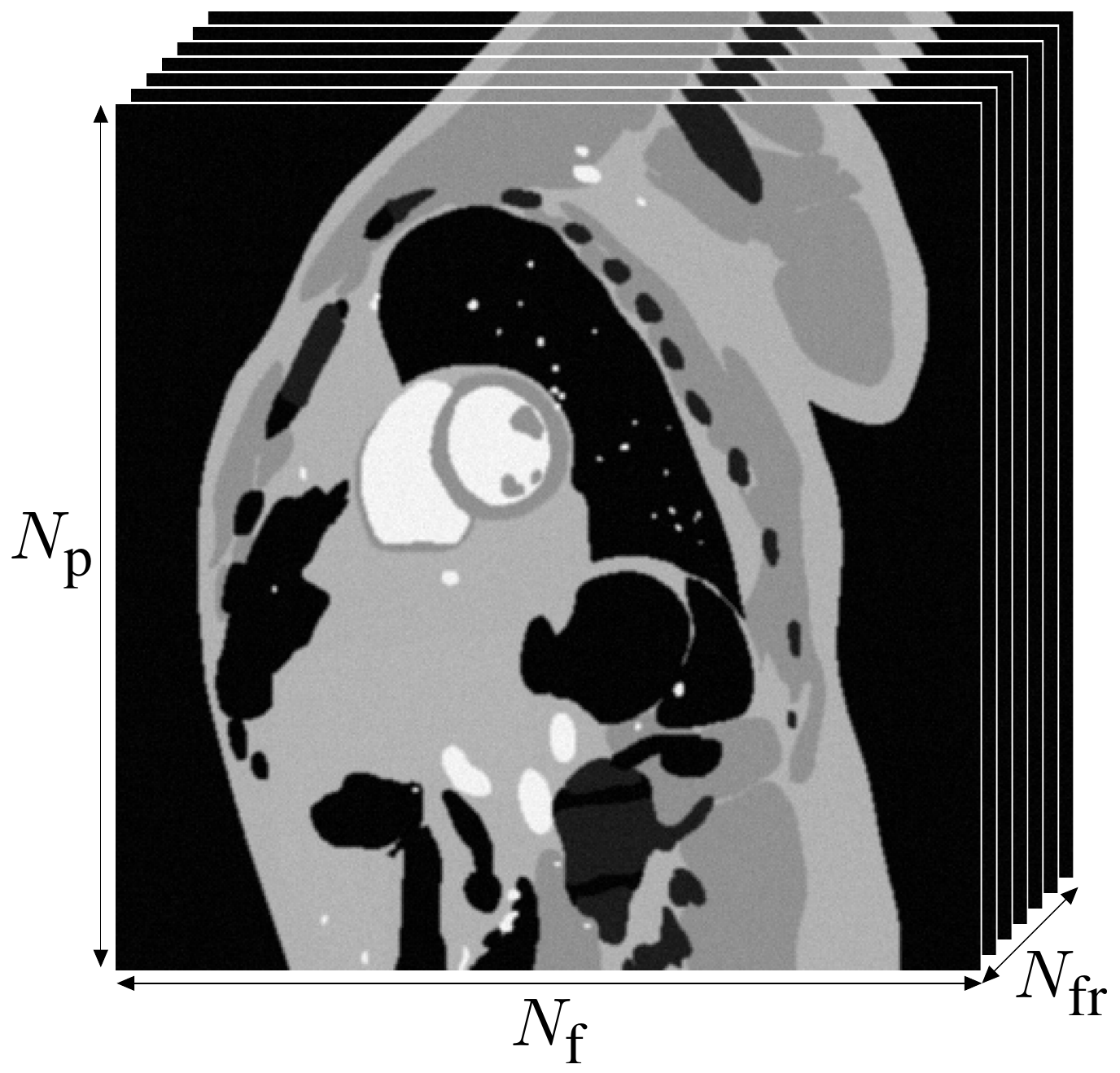}}
  \caption{(a) Complex-valued $N_{\text{f}} \times N_{\text{p}} \times N_{\text{fr}}$
    (k,t)-space dMRI data. The marked $N_{\text{f}} \times \upsilon \times N_{\text{fr}}$ box
    corresponds to the location of the faithful data, also known as ``navigator/pilot'' data in
    dMRI. (b) 1D Cartesian and (c) radial sampling trajectories in k-space. (d) Complex-valued
    $N_{\text{f}} \times N_{\text{p}} \times N_{\text{fr}}$ image-domain data.}
  \label{fig:data.description}
\end{figure*}

DMRI data take the form of a complex-valued ($\Complex$ is the set of all complex numbers)
three-way tensor $\mathbfscr{Y}$ defined on the
$(N_{\text{f}} \times N_{\text{p}} \times N_\text{fr})$-sized ``(k,t)-space''
(Fig.~\ref{fig:dmri.ktspace}), with $N_{\text{f}}, N_{\text{p}}, N_\text{fr} \in \IntegerPP$
($\IntegerPP$ is the set of all positive integers). In the seldom case where the k-space is
``fully sampled,'' the ``slice/frame'' $\mathbfcal{Y}_t$ of $\mathbfscr{Y}$ ($t$ denotes
discrete time with $t \in \{1, \ldots, N_\text{fr}\}$) collects the
$(N_{\text{f}} \times N_{\text{p}})$-sized ``k-space'' measurements at $t$. In practice, it is often the
case that the k-space data is heavily under-sampled due to physical
limitations~\cite{liang1994efficient}. Popular sampling strategies are 1-D Cartesian
(Fig.~\ref{fig:cartesian.sampling}) and radial (Fig.~\ref{fig:radial.sampling})
sampling. Sampling is denoted by the entry-wise \textit{sampling mapping}\/
$\mathscr{S}(\cdot): \mathbfcal{Y}_t \mapsto \mathscr{S} (\mathbfcal{Y}_t)$, which nullifies
the entry of $\mathbfcal{Y}_t$ when that entry is missing, while retains the entry when that
entry is successfully collected. Integers $N_{\text{f}}$ and $N_{\text{p}}$ denote the numbers
of frequency- and phase-encoding lines, respectively~\cite{zhi2000principles}, while
$N_{\text{k}} \coloneqq N_{\text{f}} N_{\text{p}}$ represents the number of entries of each
k-space frame. Typically, k-space is considered as the ``frequency domain'' of the
``image-data'' domain (Fig.~\ref{fig:dmri.image}), so that
$\mathbfcal{X}_t \coloneqq \mathscr{F}^{-1}(\mathbfcal{Y}_t)$, where $\mathscr{F}^{-1}(\cdot)$
is the 2D inverse DFT~\cite{zhi2000principles}. For convenience, the columns of
$\mathbfcal{Y}_t$ are stacked one below the other to create a single $N_{\text{k}} \times 1$
vector $\vect{y}_t \coloneqq \text{vec}(\mathbfcal{Y}_t)$ and
$\vect{Y} \coloneqq [ \vect{y}_1, \ldots, \vect{y}_{N_\text{fr}} ] \in \Complex^{ N_{\text{k}}
  \times N_{\text{fr}} }$.

It is often the case in an imputation framework for a subset of the sampled data to be
considered \textit{faithful.} In the present context, the ``low-frequency'' region, \ie, the
central region of k-space, constitutes the faithful data from which geometric information will
be extracted~\cite{shetty2020bilmdm, slavakis2022krim}. These data will be called
``navigator/pilot data.'' The navigator data of the frame $\mathbfcal{Y}_t$
(Fig.~\ref{fig:dmri.ktspace}) are gathered into the $\nu \times 1$ vector
$\check{\vect{y}}^{\text{f}}_t$, where $\nu \coloneqq \upsilon N_\text{f}$. All of these
vectors are finally stacked into the columns of
$\check{\vect{Y}}_{\text{f}} \coloneqq [ \check{\vect{y}}_1^{\text{f}}, \ldots,
\check{\vect{y}}_{N_\text{fr}}^{\text{f}}] \in \Complex^{\nu \times N_\text{fr}}$.

To offer an algorithmic scheme with manageable computational complexity as the cardinality of
the point-cloud $\{ \check{\vect{y}}^{\text{f}}_t \}_{t=1}^{N_{\text{fr}}}$ grows for datasets
with large $N_{\text{fr}}$, a subset $\{ \mathbfit{l}_k \}_{k=1}^{N_{\mathit{l}}}$, coined
\textit{landmark/representative}\/ points, with $N_{\mathit{l}} \leq N_{\text{fr}}$, is
selected from $\{ \check{\vect{y}}^{\text{f}}_t \}_{t=1}^{N_{\text{fr}}}$. Any selection
strategy can be used to identify $\{ \mathbfit{l}_k \}_{k=1}^{N_{\mathit{l}}}$. Here, the
min-max-distance strategy of~\cite{de2004sparse} is adopted along the lines
of~\cite{shetty2020bilmdm, slavakis2022krim}. For convenience, let the
$\nu \times N_{\mathit{l}}$ matrix
$\vect{L} \coloneqq [ \mathbfit{l}_1, \mathbfit{l}_2, \ldots, \mathbfit{l}_{N_{\mathit{l}}}]$.



\section{Data Modeling}

\begin{figure}[ht]
  \centering
  \includegraphics[width=0.7\linewidth]{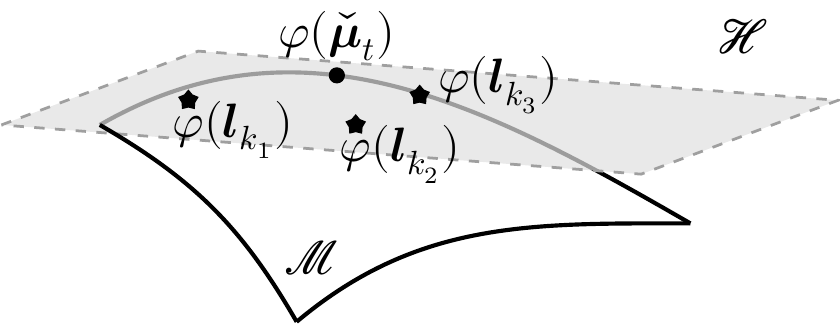}
  \caption{A ``collaborative-filtering'' approach: Points
    $\Set{ \varphi (\mathbfit{l}_{k_j}) }_{j=1}^3$, which lie into or close to the
    unknown-to-the-user manifold $\mathscr{M} \subset \mathscr{H}$, collaborate affinely to
    approximate $\varphi (\check{\bm{\mu}}_{t})$. All affine combinations of
    $\Set{ \varphi (\mathbfit{l}_{k_j})}_{j=1}^3$ define the approximating ``linear patch''
    (gray-colored plane), which mimics the concept of a tangent space to $\mathscr{M}$.}
  \label{fig:manifold.kernel.space}
\end{figure}

With $\varphi(\cdot)$ denoting the feature mapping which maps vector $\mathbfit{l}_k$ to vector
$\varphi(\mathbfit{l}_k)$ in the feature space $\mathscr{H}$, the crux of the modeling approach
is that $\{ \varphi(\mathbfit{l}_k) \}_{k=1}^{N_{\mathit{l}}}$ lie into or close to an
unknown-to-the-user smooth manifold $\mathscr{M}$~\cite{RobbinSalamon:22} embedded in
$\mathscr{H}$; see Fig.~\ref{fig:manifold.kernel.space}. To provide structured solutions, it is
assumed that $\mathscr{H}$ is a reproducing kernel Hilbert space (RKHS), equipped with a
reproducing kernel $\kappa(\cdot, \cdot): \Complex^{\nu}\times \Complex^{\nu} \to \Complex$,
with well-documented merits in approximation theory~\cite{aronszajn1950theory}. To this end,
the feature mapping $\varphi(\cdot)$ is defined as the mapping induced by the kernel $\kappa$
of $\mathscr{H}$; \cf~\cite{slavakis2022krim}. A complex-valued $\mathscr{H}$ is considered
here~\cite{bouboulis2010extension, slavakis2022krim}.


The $(i,t)$th entry $y_{it}$ of data $\vect{Y}$ is approximated as
$y_{it} \approx f_i( \check{\bm{\mu}}_{t} )$, where $f_i(\cdot): \Complex^{\nu} \to \Complex$
is an unknown non-linear function that belongs to the functional space $\mathscr{H}$, and
$\check{\bm{\mu}}_{t}$ is an unknown complex-valued $\nu \times 1$ vector. The well-known
reproducing property~\cite{aronszajn1950theory} of the RKHS $\mathscr{H}$ suggests that
$y_{it} \approx f_i( \check{\bm{\mu}}_{t} ) = \innerp{f_i}{ \varphi(
  \check{\bm{\mu}}_{t})}_{\mathscr{H}}$, where $\innerp{\cdot}{\cdot}_{\mathscr{H}}$ stands for
the inner product of $\mathscr{H}$.

The concept of tangent spaces~\cite{RobbinSalamon:22} (``linear patches'' in simple words) to
manifolds will be used to model both $f_i$ and $\varphi( \check{\bm{\mu}}_{t})$. Specifically,
it is assumed that
\begin{enumerate*}[label=\textbf{(\roman*)}]

\item $f_i$ belongs to the linear span of
  $\Set{ \varphi( \mathbfit{l}_k) }_{k=1}^{ N_{\mathit{l}} }$, \ie, there exists
  $\vect{d}_i \coloneqq [ d_{i1}, \ldots, d_{iN_{\mathit{l}}}]^{\intercal} \in
  \Complex^{N_{\mathit{l}}}$ ($\intercal$ denotes vector/matrix transposition) s.t.\
  $f_i = \sum_{k=1}^{N_{\mathit{l}}} d_{ik} \varphi( \mathbfit{l}_k ) = \bm{\Phi}( \vect{L} )
  \vect{d}_i$, where
  $\bm{\Phi}( \vect{L} ) \coloneqq [ \varphi(\mathbfit{l}_1), \ldots, \varphi(
  \mathbfit{l}_{N_{\mathit{l}}}) ]$; and

\item $\varphi( \check{\bm{\mu}}_{t})$ lies into or close to $\mathscr{M}$ and is approximated
  \textit{affinely}\/ by only a few members of
  $\Set{ \varphi( \mathbfit{l}_k) }_{k=1}^{ N_{\mathit{l}} }$, \ie, there exists a sparse
  $N_{\mathit{l}} \times 1$ vector $\vect{b}_t$ s.t.\
  $\varphi( \check{\bm{\mu}}_{t}) = \bm{\Phi}(\vect{L}) \vect{b}_t$, under the affine
  constraint $\vect{1}_{N_{\mathit{l}}}^{{\hermconj}} \vect{b}_t = 1$, where
  $\vect{1}_{N_{\mathit{l}}}$ is the $N_{\mathit{l}} \times 1$ all-one vector and $\hermconj$
  denotes complex conjugate vector/matrix transposition; \cf
  Fig.~\ref{fig:manifold.kernel.space}.

\end{enumerate*}

In other words,
$y_{it} \approx f_i( \check{\bm{\mu}}_{t} ) = \innerp{f_i}{ \varphi(
  \check{\bm{\mu}}_{t})}_{\mathscr{H}} = \innerp{\bm{\Phi}( \vect{L} ) \vect{d}_i}
{\bm{\Phi}(\vect{L}) \vect{b}_t}_{\mathscr{H}} = \vect{d}_i^{\hermconj} \vect{K} \vect{b}_t$,
where $\vect{K}$ is the complex-valued $N_{\mathit{l}} \times N_{\mathit{l}}$ matrix whose
$(k, k^{\prime})$th entry is equal to
$\innerp{\varphi(\mathbfit{l}_k)} {\varphi(\mathbfit{l}_{k^{\prime}})}_{\mathscr{H}} =
\kappa(\mathbfit{l}_k, \mathbfit{l}_{k^{\prime}})$. To offer compact notations, if
$\vect{D} \coloneqq [ \vect{d}_1, \ldots, \vect{d}_{N_{\text{k}}} ]^{{\hermconj}} \in
\Complex^{ N_{\text{k}} \times N_{\mathit{l}}}$ and
$\vect{B} \coloneqq [ \vect{b}_1, \ldots, \vect{b}_{N_{\text{fr}}} ] \in \Complex^{
  N_{\mathit{l}} \times N_{\text{fr}} }$, then data are modeled as
$\vect{Y} \approx \vect{D} \vect{K} \vect{B}$.

Choosing $\kappa(\cdot, \cdot)$ to define $\vect{K}$ entails the cumbersome tasks of cross
validation and fine tuning via extensive experimentation on data sets. A popular way to
surmount such cumbersome tasks, followed also in \cite{slavakis2022krim}, is via multiple
kernels: $\vect{Y} \approx \sum_{m=1}^M \vect{D}_m \vect{K}_m \vect{B}_m$, with a dictionary of
user-defined reproducing kernels $\Set{ \kappa_m(\cdot, \cdot) }_{m=1}^M$, and thus kernel
matrices $\Set{ \vect{K}_m }_{m=1}^M$, complex-valued $N_{\text{k}} \times N_{\mathit{l}}$
matrices $\Set{ \vect{D}_m }_{m=1}^M$, and $N_{\mathit{l}} \times N_{\text{fr}}$ sparse
matrices $\Set{ \vect{B}_m }_{m=1}^M$ satisfying
$\vect{1}^{{\hermconj}}_{N_{\mathit{l}}} \vect{B}_m = \vect{1}_{N_{\text{fr}}}^{{\hermconj}}$.



To reduce the computational burden and effect low-rank constraints in the resultant inverse
problem, each $N_{\mathit{l}} \times N_{\mathit{l}}$ matrix $\vect{K}_m$ is substituted
in~\cite{slavakis2022krim} by its low-dimensional $d \times N_{\mathit{l}}$ rendition
$\check{\vect{K}}_m$ in
$\vect{Y} \approx \sum_{m=1}^M \check{\vect{D}}_m \check{\vect{K}}_m \vect{B}_m$, with
$d\ll N_{\mathit{l}}$ and where $\Set{ \check{\vect{D}}_m }_{m=1}^M$ are low-rank
$N_{\text{k}} \times d$ matrices. To this end, $\check{\vect{K}}_m$ was computed from
$\vect{K}_m$ via a dimensionality-reduction module (pre-step)
in~\cite{slavakis2022krim}. However, such a dimensionality-reduction pre-step introduces the
following drawbacks:
\begin{enumerate*}[label=\textbf{(\roman*)}]

\item when the numbers $M$ of kernels and $N_{\mathit{l}}$ of landmark points are large, the
  dimensionality-reduction module inflicts heavy computations, while fine-tuning its
  hyperparameters becomes a labor-intensive task; and

\item the error from compressing $\vect{K}_m$ into $\check{\vect{K}}_m$ may propagate to the
  next phase in the KRIM framework.

\end{enumerate*}

MultiL-KRIM avoids the previous drawbacks as follows:
\begin{align}
  \vect{Y} \approx \sum\nolimits_{m=1}^M \vect{D}_m^{(1)} \vect{D}_m^{(2)} \cdots
  \vect{D}_m^{(Q)} \vect{K}_m \vect{B}_m \,, \label{multi.kernel.nodimred}
\end{align}
where $\vect{D}_m^{(q)} \in \Complex^{d_{q-1}\times d_q}$, the inner matrix dimensions
$\Set{ d_q }_{q=1}^{Q-1}$ are user-defined, with $d_0 \coloneqq N_{\text{k}}$ and
$d_Q \coloneqq N_{\mathit{l}}$. Notice that for $Q=2$, the term $\vect{D}_m^{(2)} \vect{K}_m$
may be considered as the dimensionality-reduced $\check{\vect{K}}_m$
in~\cite{slavakis2022krim}. Nonetheless, $\Set{\vect{D}_m^{(q)}}_{(q, m)}$ are identified
during a single-stage learning task, avoiding any pre-steps with their hyperparameter tuning
and errors. Note that $Q>1$ in \eqref{multi.kernel.nodimred} may offer considerable savings in
computations with respect to the $Q = 1$ case. Indeed, the number of unknowns that need to be
identified in \eqref{multi.kernel.nodimred} for $Q > 1$ is
$N_{Q>1} = M ( \sum_{q=1}^Q d_{q-1}d_q + N_{\text{fr}}N_{\mathit{l}} )$, as opposed to
$N_{Q=1} = M ( N_{\text{k}}N_{\mathit{l}} + N_{\text{fr}}N_{\mathit{l}} )$ in the $Q = 1$
case. If $N_{\mathit{l}}$ is large, then $\Set{ d_q }_{q=1}^{Q-1}$ can be chosen so that
$N_{Q>1} \ll N_{Q=1}$.





\section{Inverse Problem And its Iterative Solution}



Letting
$\vectcal{D}_1 \coloneqq [\vect{D}_1^{(1)}, \vect{D}_2^{(1)}, \ldots, \vect{D}_M^{(1)}] \in
\Complex^{N_{\text{k}} \times d_1M}$,
$\vectcal{B} \coloneqq [\vect{B}_1^{\hermconj}, \vect{B}_2^{\hermconj}, \ldots,
\vect{B}_{M}^{\hermconj}]^{\hermconj} \in \Complex^{MN_{\mathit{l}} \times N_{\text{fr}}}$, and
the block diagonal matrices
$\vectcal{D}_q \coloneqq \bdiag{(\vect{D}_1^{(q)}, \vect{D}_2^{(q)}, \ldots, \vect{D}_M^{(q)})}
\in \Complex^{d_{q-1}M \times d_{q}M}$, $q \in \Set{2, \ldots, Q}$,
$\vectcal{K} \coloneqq \bdiag{(\vect{K}_1, \vect{K}_2, \ldots, \vect{K}_M)} \in
\Complex^{MN_{\mathit{l}} \times MN_{\mathit{l}}}$, \eqref{multi.kernel.nodimred} takes the
form
$\vect{Y} \approx \vectcal{D}_1 \vectcal{D}_2 \cdots \vectcal{D}_Q \vectcal{K} \vectcal{B}$.


In dMRI, it is often the case that inverse problems are formulated in the image domain, since,
after all, this is the domain where the end-product lies in. To this end, by using $\vect{X}$
to denote the image-domain data and by defining
$\vectcal{A}_1 \coloneqq \mathscr{F}^{-1}( \vectcal{D}_1 )$ (let also
$\vectcal{A}_q \coloneqq \vectcal{D}_q, \forall q \in \Set{2, \ldots, Q}$, for uniform
notations),
the following inverse problem is postulated:
\begin{subequations}\label{eq:recovery.task.general}
  \begin{align}
    \min_{ (\vect{X}, \vect{Z}, \Set{\vectcal{A}_q}_{q=1}^Q, \vectcal{B}) }
    {}\ & {} \tfrac{1}{2} \norm{ \vect{X} - \vectcal{A}_1 \vectcal{A}_2 \cdots
          \vectcal{A}_Q \vectcal{K} \vectcal{B} }^2_{\text{F}} \notag \\
        & + \lambda_1 \norm{\vectcal{B}}_1
          + g(\vect{X}, \vect{Z}) +
          h(\Set{\vectcal{A}_q}_{q=1}^Q) \label{eq:recovery.task.loss} \\
    \text{s.to} {}\
        & \mathscr{S}(\vect{Y}) = \mathscr{S}
          \mathscr{F}(\vect{X})\,, \label{recovery.task.consistency} \\
        & \vect{1}_{N_{\mathit{l}}}^{{\hermconj}} \vect{B}_m =
          \vect{1}_{N_{\text{fr}}}^{{\hermconj}}\,,
          \forall m \in \Set{1, \ldots, M} \label{recovery.task.B}\,, \\
 & \vectcal{A}_q \text{is block diagonal}, \forall q\in \Set{2, \ldots,
   Q}\,, \label{recovery.task.A}
  \end{align}
\end{subequations}
where \eqref{recovery.task.consistency} enforces consistency of the desired $\vect{X}$ with the
data collected in the (k,t) domain, $\norm{\cdot}_1$ in \eqref{eq:recovery.task.loss} is used
to impose sparsity on $\vectcal{B}$, and the convex regularizing functions $g(\cdot)$ and
$h(\cdot)$ are used to incorporate prior knowledge. More specifically, $g(\cdot)$ employs not
only variable $\vect{X}$ but also the auxiliary $\vect{Z}$ to facilitate computations. For
example, in the case where the dMRI data capture a periodic organ movement over a static
background, then
$g(\vect{X}, \vect{Z}) \coloneqq (\lambda_2/2)\norm{\vect{Z} - \mathscr{F}_{\text{t}}
  (\vect{X})}_{\text{F}}^2 + \lambda_3 \norm{\vect{Z}}_1$, where
$\mathscr{F}_{\text{t}}(\cdot)$ stands for the temporal 1D DFT operator which acts on rows of
the matrix $\vect{X}$. Note that a row of $\vect{X}$ corresponds to the time series, of
length $N_{\text{fr}}$, of a single pixel in the image domain. Additionally, the designer can
choose
$h(\Set{\vectcal{A}_q}_{q=1}^Q) \coloneqq (\lambda_4/2) \sum_{q=1}^Q
\norm{\vectcal{A}_q}_{\text{F}}^2$ to avoid unbounded solutions which may appear due to
$\vectcal{A}_1 \vectcal{A}_2 \cdots \vectcal{A}_Q$ in \eqref{eq:recovery.task.general}.


\begin{algorithm}[H]
  \caption{Solving MultiL-KRIM's inverse problem}\label{alg:recovery.k}
  \begin{algorithmic}[1]


    \ENSURE Limit point $\hat{\vect{X}}^{(*)}$ of sequence
    $(\hat{\vect{X}}^{(n)})_{n\in\IntegerP}$.



    \STATE Fix $\hat{\mathbfcal{S}}^{(0)}$, $\gamma_0\in (0,1]$, and $\zeta\in (0,1)$.

    \WHILE{$n\geq 0$} \label{alg.step:resume.k}

    \STATE { Available are
      $\hat{\vectcal{S}}^{(n)}$ and $\gamma_n$.  }

    \STATE {$\gamma_{n+1} \coloneqq \gamma_n (1 - \zeta \gamma_n)$.}

    \STATE Solve sub-tasks \eqref{eq:recovery.subtasks}.


    \STATE
    {$\hat{\vectcal{S}}^{(n+1)} \coloneqq \gamma_{n+1} \hat{\vectcal{S}}^{(n+1/2)} +
      (1-\gamma_{n+1}) \hat{\vectcal{S}}^{(n)}$.}


    \STATE {Set $n \leftarrow n+1$ and go to step~\ref{alg.step:resume.k}.}

    \ENDWHILE
  \end{algorithmic}
  \label{alg1}
\end{algorithm}

Alg.~\ref{alg1} sketches a solution to \eqref{eq:recovery.task.general}, where
$\hat{\mathbfcal{S}}^{(n+k/2)} \coloneqq ( \hat{\vect{X}}^{(n+k/2)}, \hat{\vect{Z}}^{(n+k/2)},
\hat{\vectcal{A}}_1^{(n+k/2)}, \ldots, \hat{\vectcal{A}}_Q^{(n+k/2)},
\hat{\vectcal{B}}^{(n+k/2)} )$, $\forall n\in\IntegerP$, $\forall
k\in\Set{0,1}$. Alg.~\ref{alg1} is based on the successive-convex-approximation framework
of~\cite{facchinei2015parallel}, which guarantees convergence to a stationary point of the
loss. The following convex sub-tasks need to be solved per iteration:
$\forall q\in \Set{1, \ldots, Q}$,
\begin{subequations}\label{eq:recovery.subtasks}
  \begin{alignat}{3}
    && \hat{\vect{X}}^{(n + 1/2) }
    && \in \arg\min_{ {\vect{X}} } {}
    & {}\ \tfrac{1}{2} \norm{{\vect{X}} - \hat{\vectcal{A}}_1^{(n)} \cdots
      \hat{\vectcal{A}}_Q^{(n)} \vectcal{K} \hat{\vectcal{B}}^{(n)} }_{\text{F}}^2 \notag \\
    &&&&& {}\ + \tfrac{\lambda_2}{2} \norm{ \hat{\vect{Z}}^{(n)} -
          \mathscr{F}_{\text{t}}( {\vect{X}} ) }_{\text{F}}^2 + \tfrac{\tau_X}{2} \norm{
          \vect{X} - \hat{\vect{X}}^{(n) }}_{\text{F}}^2 \notag \\
    &&&& \text{s.to} {} & {}\ \mathscr{S}(\vect{Y}) = \mathscr{S} \mathscr{F}( {\vect{X}} ) \,,
                          \label{eq:task.min.X} \\
    && \hat{\vect{Z}}^{(n + 1/2) }
    && \in \arg\min_{\vect{Z}} {}
    & {}\ \tfrac{\lambda_2}{2} \norm{ \vect{Z} - \mathscr{F}_{\text{t}}( \hat{\vect{X}}^{(n)})
      }^2_{\text{F}} + \lambda_3 \norm{ \vect{Z} }_1 \notag \\
    &&&&& {}\ + \tfrac{\tau_Z}{2} \norm{ \vect{Z} - \hat{\vect{Z}}^{(n)
          }}_{\text{F}}^2 \,, \label{eq:task.Z}\\
    && \hat{\vectcal{A}}_q^{(n + 1/2)} {}
    && {} \in \arg\min_{\vectcal{A}_q} {}
    & {}\ \tfrac{1}{2} \norm{ \hat{\vect{X}}^{(n)} - \hat{\vectcal{A}}_1^{(n)} \cdots
      \vectcal{A}_q \cdots \hat{\vectcal{A}}_Q^{(n)} \vectcal{K}
      \hat{\vectcal{B}}^{(n)} }_{\text{F}}^2 \notag \\
    &&&& {} & {}\ + \tfrac{\lambda_4}{2} \norm{ \vectcal{A}_q }_{\text{F}}^2
              + \tfrac{\tau_A}{2} \norm{ \vectcal{A}_q - \hat{\vectcal{A}}_q^{(n)}
              }_{\text{F}}^2 \notag \\
    &&&& \text{s.to} {} & {}\ \vectcal{A}_q\ \text{is block-diagonal}\,, \forall q\in \Set{2,
                          \ldots, Q}\,, \label{eq:task.min.A} \\
    && \hat{\vectcal{B}}^{(n + 1/2)}
    && \in \arg \min_{\vectcal{B}} {}
    & {}\ \tfrac{1}{2} \norm{ \hat{\vect{X}}^{(n)}- \hat{\vectcal{A}}_1^{(n)} \cdots
      \hat{\vectcal{A}}_Q^{(n)} \vectcal{K} \vectcal{B} }_{\text{F}}^2 \notag \\
    &&&&& {}\ + \lambda_1 \norm{ \vectcal{B} }_1
          + \tfrac{\tau_B}{2} \norm{ \vectcal{B} - \hat{\vectcal{B}}^{(n)} }_{\text{F}}^2
          \notag \\
    &&&& \text{s.to} {}
    & {}\ \vect{1}_{N_{\mathit{l}}}^{{\hermconj}} \vect{B}_m =
      \vect{1}_{N_\text{fr}}^{{\hermconj}} \,, \forall m \in \Set{1, \ldots, M}
      \,. \label{eq:task.min.B}
  \end{alignat}
\end{subequations}
Sub-task \eqref{eq:task.min.B} is a composite convex minimization task under affine
constraints, hence it can be solved by \cite{slavakis2018fejer}, while \eqref{eq:task.min.X},
\eqref{eq:task.Z} and \eqref{eq:task.min.A} have closed form
solutions~\cite{slavakis2022krim}. More specifically, the unique solution to \eqref{eq:task.Z}
is provided by the well-known soft-thresholding operator~\cite{slavakis2022krim}.



\section{Numerical Tests}\label{sec:numerical}

\begin{figure}[t!]
  \centering
  \subfloat[\label{fig:plot.cartesian}]
  {\includegraphics[width =
    .8\columnwidth]{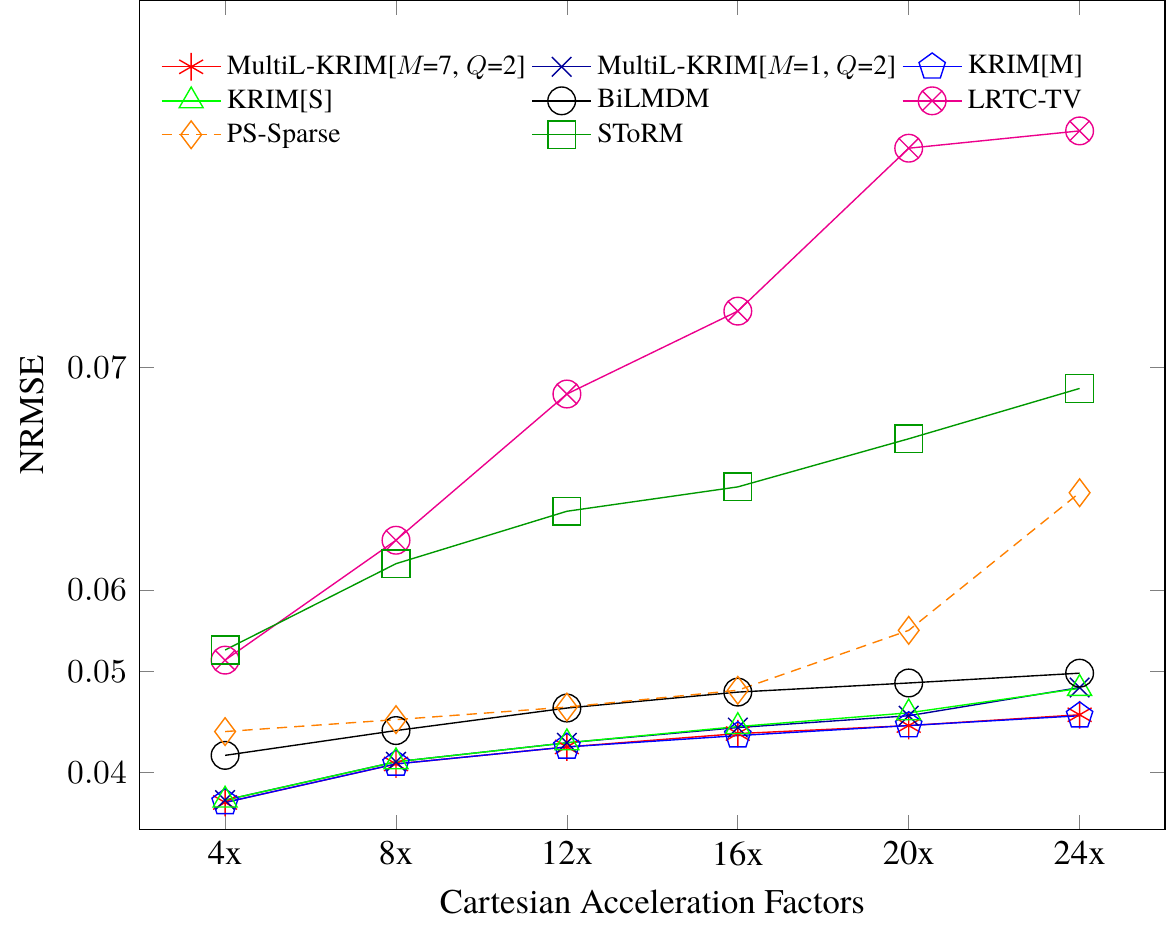}} \\
    \subfloat[\label{fig:plot.radial}]
  {\includegraphics[width = .8\columnwidth]{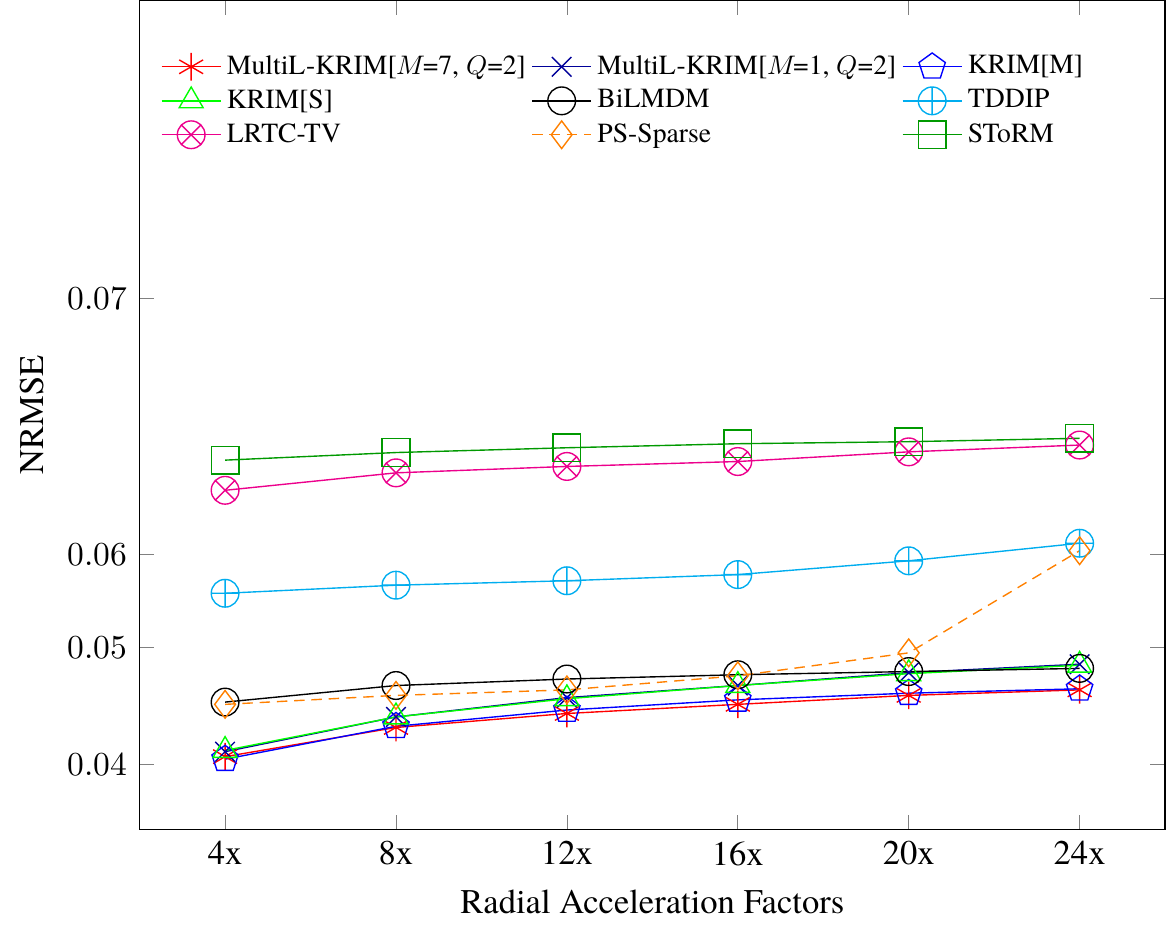}}
  \centering\caption{NRMSE values
    at different Cartesian \protect\subref{fig:plot.cartesian} and radial
    \protect\subref{fig:plot.radial} acceleration rates. The NRMSE values are plotted on a
    log-axis.}
  \label{fig:nrmse.v.acceleration}
\end{figure}


Following~\cite{zhao2012image, nakarmi2017m, nakarmi2018mls, shetty2020bilmdm}, the proposed
framework is validated on the magnetic resonance extended cardiac-torso (MRXCAT) cine phantom
dataset~\cite{wissmann2014mrxcat} under both radial and Cartesian sampling. MultiL-KRIM[$M,Q$]
is compared against its predecessors KRIM~\cite{slavakis2022krim} and
BiLMDM~\cite{shetty2020bilmdm}, as well as against the low-rank tensor factorization with total
variation (LRTC-TV)~\cite{Li_Ye_Xu_2017} (designed also for dMRI data), the DIP-based
TDDIP~\cite{yoo2021time} (designed especially for radial sampling), and the popular
PS-Sparse~\cite{zhao2012image} and SToRM~\cite{poddar2016dynamic}. Tags KRIM[S] and KRIM[M]
refer to \cite{slavakis2022krim} for the case of a single ($M=1$) and multiple ($M>1$) kernels,
respectively. Comparisons of KRIM and BiLMDM against several other state-of-the-art methods on
the same data can be found in~\cite{slavakis2022krim, shetty2020bilmdm}. All methods were
finely tuned to achieve best performance.

Parameter $Q \in \Set{2, 3, 4}$, and for each $Q$ the inner matrix dimensions are set as
follows:
\begin{enumerate*}[label=\textbf{(\roman*)}]

\item if $Q=2$, then $d_1 = 6$;
\item if $Q=3$, then $(d_1, d_2) = (2, 6)$; and
\item if $Q=4$, then $(d_1, d_2, d_3) = (2, 4, 6)$.

\end{enumerate*}
The inner dimension parameter of KRIM is $d \coloneqq 6$. Parameter $M \in \Set{1, 7}$ for both
MultiL-KRIM and KRIM, with choices of kernels as in~\cite{slavakis2022krim}. Number of landmark
points is $N_{\mathit{l}} \coloneqq 100$. Since \eqref{eq:recovery.task.general} is a
non-convex task, and due to the well-known fact that the limit point of any iterative algorithm
which seeks a stationary point of \eqref{eq:recovery.task.general} depends on the starting
point $\hat{\mathbfcal{S}}^{(0)}$, Alg.~\ref{alg1} was run multiple times for each scenario,
with different $\hat{\mathbfcal{S}}^{(0)}$ per run, and all reported metric values are the mean
values of all those multiple runs. The software code for (MultiL-)KRIM and BiLMDM was written
in Julia~\cite{bezanson2017julia}. All tests were run on an 8-core Intel(R) i7-11700 2.50GHz
CPU with 32GB RAM.


The main evaluation metric is the normalized root mean square error
$\text{NRMSE} \coloneqq \norm{ \vect{X} -\hat{\vect{X}} }_{\text{F}} / \norm{ \vect{X}
}_{\text{F}}$, where $\vect{X}$ denotes the image data obtained from fully sampled
(k,t)-space data, and $\hat{\vect{X}}$ represents the estimate of $\vect{X}$. Additionally,
reconstructions of high-frequency regions are evaluated by the high-frequency error norm (HFEN)
and two sharpness measures M1 (intensity-variance based) and M2 (energy of the image
gradient)~\cite[(43) and (46)]{subbarao1993focusing}.
Lastly, the structural similarity measure (SSIM)~\cite{wang2004image} captures local
similarities in pixel intensities after normalizing for luminance and contrast.


It can be seen in Fig.~\ref{fig:nrmse.v.acceleration}, and more precisely in
Tables~\ref{tab:mrxcat.cartesian} and~\ref{tab:mrxcat.radial}, that MultiL-KRIM matches the
performance of KRIM. Nevertheless, MultiL-KRIM shows remarkable reduction in computational time
by up to 65\% for multiple kernels and 30\% for a single kernel. To showcase
that the reduction in computational time is because of the novel factorization approach, the
reported times of KRIM[M] and KRIM[S] \textit{do not}\/ include the time of KRIM's
dimensionality-reduction pre-step. The tests also demonstrate the better performance of
MultiL-KRIM over the state-of-the-art LRTC-TV and TDDIP. Notice that TDDIP takes much longer
time to run on the CPU due to its deep CNN architecture. On the other hand, LRTC-TV
deteriorates quickly with high acceleration rates under Cartesian sampling, with a high
computational footprint.


\begin{table}[ht!]
  \renewcommand*{\arraystretch}{1.2}
  \caption{Performance on Cartesian sampling (acceleration rate: 20x)}
  \centering \resizebox{\columnwidth}{!}{%
    {\begin{tabular}{|c|l|l|l|l|l|l|} \hline Methods $\setminus$ Metrics & \multicolumn{1}{c|}{\textbf{NRMSE}}
       & \multicolumn{1}{c|}{\textbf{SSIM}} & \multicolumn{1}{c|}{\textbf{HFEN}}
       & \multicolumn{1}{c|}{\textbf{M1}} & \multicolumn{1}{c|}{\textbf{M2}}
       & \multicolumn{1}{c|}{\textbf{Time}} \\  \hline
       \rowcolor{tableMulti}
       \textbf{MultiL-KRIM[$M=7, Q=2$]} & ${\mathbf{0.0443}}$ & $\mathbf{0.8698}$ & ${0.1147}$
       & $37.26$ & $\mathbf{1.4 \times 10^6}$ & 1.2hrs \\ \hline
       \rowcolor{tableSingle}
       \textbf{MultiL-KRIM[$M=1, Q=2$]} & ${0.0453}$ & ${0.8686}$ & ${0.1269}$ & $37.24$
                                          & ${1.3 \times 10^6}$ & 46min
       \\  \hline
       \rowcolor{tableMulti}
       \textbf{MultiL-KRIM[$M=7, Q=3$]} & ${{0.0444}}$ & ${0.8695}$ & ${0.1149}$ & $37.26$
                                          & $\mathbf{1.4 \times 10^6}$ & 1.3hrs
       \\ \hline
       \rowcolor{tableSingle}
       \textbf{MultiL-KRIM[$M=1, Q=3$]} & ${0.0454}$ & ${0.8684}$ & ${0.1270}$ & $37.24$
                                          & ${1.3 \times 10^6}$ & 42min
       \\  \hline
       \rowcolor{tableMulti}
       \textbf{MultiL-KRIM[$M=7, Q=4$]} & ${\mathbf{0.0443}}$ & ${0.8696}$ & ${0.1147}$ & $37.26$
                                          & $\mathbf{1.4 \times 10^6}$ & 1.4hrs
       \\ \hline
       \rowcolor{tableSingle}
       \textbf{MultiL-KRIM[$M=1, Q=4$]} & ${0.0453}$ & ${0.8686}$ & ${0.1269}$ & $37.24$
                                          & ${1.3 \times 10^6}$ & 43min
       \\ \hline
       \rowcolor{tableMulti}
       \textbf{KRIM[M]}~\cite{slavakis2022krim} & $\mathbf{0.0443}$ & ${0.8696}$ & $\mathbf{0.1136}$ & $\mathbf{37.34}$
                                          & $\mathbf{1.4 \times 10^6}$ & 3.4hrs \\ \hline
        \rowcolor{tableSingle}
       \textbf{KRIM[S]}~\cite{slavakis2022krim} & $0.0450$ & $0.8670$ & $0.1149$
       & $37.25$ & $1.3 \times 10^6$ & 55min \\
          \hline \textbf{BiLMDM}~\cite{shetty2020bilmdm} & $0.0488$ & $0.8589$ & $0.1423$ & $37.08$ & $1.3 \times 10^6$
& 40min \\
          \hline \textbf{LRTC-TV}~\cite{Li_Ye_Xu_2017} & $0.1645$ & $0.5942$ & $0.3834$ & $31.52$
& ${8.9 \times 10^5}$ & 8.3hrs \\
          \hline \textbf{SToRM}~\cite{poddar2016dynamic} & $0.0850$ & $0.8110$ & $0.2504$ & $37.02$ &
          $1.2 \times 10^6$ & 58min \\
          \hline \textbf{PS-Sparse}~\cite{zhao2012image} & $0.0550$ & $0.8198$ & $0.1483$ &
          $37.18$ & $1.2 \times 10^6$ & \textbf{15min} \\ \hline
    \end{tabular}}
    }\label{tab:mrxcat.cartesian}

  \end{table}

  \begin{table}[ht!]
    \renewcommand*{\arraystretch}{1.2}
    \caption{Performance of radial sampling (acceleration rate: 16x)}
    \centering \resizebox{\columnwidth}{!}{%
      {\begin{tabular}{|c|l|l|l|l|l|l|} \hline Methods $\setminus$ Metrics & \multicolumn{1}{c|}{\textbf{NRMSE}}
         & \multicolumn{1}{c|}{\textbf{SSIM}} & \multicolumn{1}{c|}{\textbf{HFEN}}
         & \multicolumn{1}{c|}{\textbf{M1}} & \multicolumn{1}{c|}{\textbf{M2}}
         & \multicolumn{1}{c|}{\textbf{Time}} \\  \hline
         \rowcolor{tableMulti}
         \textbf{MultiL-KRIM[$M=7, Q=2$]} & ${\mathbf{0.0448}}$ & $\mathbf{0.8680}$
                                              &${\mathbf{0.1023}}$ & $37.35$ & ${1.4 \times 10^6}$ & 1.1hrs
         \\ \hline
         \rowcolor{tableSingle}
         \textbf{MultiL-KRIM[$M=1, Q=2$]} & ${0.0465}$ & ${0.8618}$ & ${0.1305}$ & $37.26$
                                            &${1.4 \times 10^6}$ & 44min
         \\  \hline
         \rowcolor{tableMulti}
         \textbf{MultiL-KRIM[$M=7, Q=3$]} & ${{0.0450}}$ & ${0.8670}$ & ${{0.1130}}$ & $37.35$
                                            & ${1.4 \times 10^6}$ & 1.2hrs
         \\ \hline
         \rowcolor{tableSingle}
         \textbf{MultiL-KRIM[$M=1, Q=3$]} & ${0.0465}$ & ${0.8618}$ & ${0.1305}$ & $37.26$
                                            & ${1.4 \times 10^6}$ & 41min
         \\  \hline
         \rowcolor{tableMulti}
         \textbf{MultiL-KRIM[$M=7, Q=4$]} & ${\mathbf{0.0448}}$ & $\mathbf{0.8680}$
                                              & ${\mathbf{0.1023}}$
         & $37.35$ & ${1.4 \times 10^6}$ & 1.2hrs \\ \hline
         \rowcolor{tableSingle}
         \textbf{MultiL-KRIM[$M=1, Q=4$]} & ${0.0465}$ & ${0.8616}$ & ${0.1307}$ & $37.26$
                                            & ${1.4 \times 10^6}$ & 42min
         \\ \hline
         \rowcolor{tableMulti}
         \textbf{KRIM[M]}~\cite{slavakis2022krim} & ${0.0450}$ & ${0.8670}$ & ${0.1136}$ & $37.34$
                                            & ${1.4 \times 10^6}$ & 3hrs \\ \hline
         \rowcolor{tableSingle}
         \textbf{KRIM[S]}~\cite{slavakis2022krim} & $0.0465$ & $0.8618$ & $0.1301$
         & $37.26$ & $1.4 \times 10^6$ & 58min \\
          \hline \textbf{BiLMDM}~\cite{shetty2020bilmdm} & $0.0475$ & $0.8560$ & $0.1491$ & $37.30$ & $1.4 \times 10^6$
                                                                             & 1.4hrs \\
       \hline \textbf{TDDIP}~\cite{yoo2021time} & $0.0579$ & $0.8167$ & $0.2195$ & $\mathbf{39.81}$
                                          & $1.2 \times 10^6$ & 20hrs \\
       \hline \textbf{LRTC-TV}~\cite{Li_Ye_Xu_2017} & $0.0738$ & $0.8063$ & $0.3725$ & $37.06$
                                          & $\mathbf{1.7 \times 10^6}$ & 8.6hrs \\
          \hline \textbf{SToRM}~\cite{poddar2016dynamic} & $0.0753$ & $0.8319$ & $0.3694$ & $37.38$ &
          $1.6 \times 10^6$ & 30min \\
          \hline \textbf{PS-Sparse}~\cite{zhao2012image} & $0.0496$ & $0.7908$ & $0.1733$ &
          $37.31$ & $1.4 \times 10^6$ & \textbf{15min} \\ \hline
    \end{tabular}}
    }\label{tab:mrxcat.radial}
\end{table}



\section{Conclusions}

This paper extends the KRIM framework~\cite{slavakis2022krim} into a faster kernelized matrix
factorization framework which avoids KRIM's dimensionality reduction pre-step. Numerical tests
demonstrate that the proposed data-modeling approach matches the reconstruction performance of
KRIM under both Cartesian and radial data sampling, but with significant reduction in
computational time, and outperforms at the same time popular methods as well as
state-of-the-art tensor-based and deep-image-prior schemes.


\clearpage
\bibliography{ref.bib}

\end{document}